\documentclass[preprint,aps]{revtex4}
\usepackage{amsmath}
\usepackage{amsfonts}
\usepackage{amssymb,amscd,amsbsy}
\usepackage{epsfig}
\usepackage{dsfont}
\usepackage{color}

\begin{document}

\title{Crystallization in suspensions of hard spheres: A Monte Carlo and Molecular Dynamics simulation study} 

\author{T. Schilling}
\affiliation{Theory of Soft Condensed Matter, Universit\'e du Luxembourg, Luxembourg, Luxembourg}
\author{S. Dorosz, H.J. Sch{\"o}pe}
\affiliation{Institut f{\"ur} Physik, Johannes Gutenberg--Universit\"at Mainz,  D--55099 Mainz, Germany}
\author{G. Opletal}
\affiliation{School of Applied Sciences, RMIT University, Melbourne Australia}

\begin{abstract}
The crystallization of a metastable melt is one of the most important non 
equilibrium phenomena in condensed matter physics, and hard sphere colloidal 
model systems have been used for several decades to investigate this process 
by experimental observation and computer simulation. Nevertheless, there is 
still an unexplained discrepancy between simulation data and experimental 
nucleation rate densities. In this paper we examine the nucleation process 
in hard spheres using molecular dynamics and Monte Carlo simulation. 
We show that the crystallization 
process is mediated by precursors of low orientational bon-order 
and that our simulation data fairly match the experimental data sets.
\end{abstract}

\maketitle

\section{Introduction}
Hard spheres are often used as a model system to study the liquid to 
crystal transition. Already more than fifty years ago the existence of the
freezing transition in hard spheres has been shown by computer simulation 
methods \cite{alder.wainwright:1957}. As the interaction potential between 
two hard spheres is infinite when they overlap and zero otherwise, the phase 
behaviour is determined only by entropy rather than by a competition between 
enthalpy and entropy. The simple interaction potential makes hard spheres 
a particularly popular model system for computer simulation studies 
of crystallization and the competing glass transition 
(see e.g.~Refs.~\cite{Truskett1998,Auer2001,Gruhn2001,OMalley2003,Schilling:2010,Zaccarelli:2009,Kawa2010,Charbonneau:2010,Filion:2010}).  

Hard sphere--like systems can also be realized experimentally in 
colloidal suspensions since the 1980's \cite{Pusey:1986}. 
Using scattering techniques as well as microscopy, the crystallization 
process and the competing glass transition have been studied in detail 
during the past decade (see e.g. 
Refs.~\cite{Kegel2000,Gasser:2001,Weeks:2002,Cheng2002,Schoepe:2006}).

The recent interest in studying the crystallization process of hard spheres 
using computer simulation has been triggered in particular by the following 
reasons:   

Crystal nucleation from a supersaturated liquid is a typical ``rare event''. 
It occurs (by definition) after an induction time that is much longer than 
the time-scale for thermalization of the microscopic degrees of freedom of 
the system, and it changes the properties of the system drastically. 
Computer simulation of rare events requires special techniques in order
to avoid wasting large amounts of CPU time on irrelevant microscopic 
fluctuations. For the past decade crystal nucleation has been commonly used 
as an example problem to apply rare event sampling techniques. 
However, in the mean-time 
computers have become fast enough to sample crystal nucleation by
``brute force'' simulation in simple model systems, such as hard spheres. 
Hence hard spheres have recently been used to test the predictions of 
rare event sampling techniques (such as e.g.~results obtained
by Umbrella Sampling \cite{Auer2001}) and to compare 
nucleation pathways directly to experiment \cite{Filion:2010, Schilling:2010}.

New experiments as well as simulations show deviations from the classical 
picture of crystallization, indicating that crystallization in hard sphere 
systems starts with the formation of precursors (low symmetry clusters, 
medium range ordered crystals) before real crystal are 
formed \cite{Schoepe:2006,Iacopini2009,Schilling:2010,Kawa2010}.  
Similar observations have been made studying crystal nucleation in atomic 
systems using dynamical density functional theory \cite{Toth:2010}. 
Furthermore it was suggested that these precursors are linked with structural 
and dynamical heterogeneities of the meta-stable melt and that the formation 
of precursors might be linked with the glass 
transition \cite{Schoepe2007,Kawasaki:2010b}. Hence the topic of 
crystallization in hard spheres is currently being revisited within 
computer simulation sutdies. 

We have recently published results on the crystallization mechanism in hard 
spheres that were obtained by a brute force MC 
simulation \cite{Schilling:2010}. 
Here we will add results on nucleation rates and a comparison of two types of 
microscopic dynamics with experimental data.

\section{Simulation Details}

\label{sec:Sim}

In order to test if
the details of the short time dynamics affect the nucleation behaviour
we have performed two types of simulation:
event driven Molecular Dynamics (Newtonian free flight and collisions) and
Monte Carlo simulations using only small translational moves (mimicking 
Brownian dynamics on longtime-scales \cite{Berthier:2007,Sanz:2010}).

In both cases we monitored crystallization by means of the q6q6-bond 
order parameter 
\cite{Steinhardt.Nelson.Ronchetti:1983,Wolde.RuizM.Frenkel:1995}, the 
definition of which we briefly recapitulate:
For a particle $i$ with $n(i)$ 
neighbours, the local orientational structure is characterized by
\[
\bar{q}_{lm}(i) := \frac{1}{n(i)}\sum_{j=1}^{n(i)} Y_{lm}\left(\vec{r}_{ij}\right)\quad ,
\]
where $ Y_{lm}\left(\vec{r}_{ij}\right)$ are the spherical harmonics 
corresponding to the orientation of the vector $\vec{r}_{ij}$ between 
particle $i$ and its neighbour $j$ in a given coordinate frame.   
We are interested in local fcc-, hcp- or rcp-structures. Therefore we 
consider $l=6$. 
A vector $\vec{q}_{6}(i)$ is asigned to each particle, the elements 
$m=-6 \dots 6$  of which are defined as 
\begin{equation}
q_{6m}(i) := \frac{\bar{q}_{6m}(i)}{\left(\sum_{m=-6}^6|\bar{q}_{6m}(i)|\right)^{1/2}} \quad . \label{Defq6q6}
\end{equation}
We counted particles counted as neighoburs if their distance 
$r_{ij} < 1.4$. Two neighbouring 
particles $i$ and $j$ were regarded as ``bonded'' within a crystalline 
region, if $\vec{q_6}(i)\cdot \vec{q_6}(j) > 0.7$. We define $n_b(i)$ as 
the number of ``bonded'' neighbours of the $i$th particle. 
If a particle has $n_b > 10$ (i.e. an almost perfectly hexagonally 
ordered surrounding), we call it ``crystalline''. A cluster of particles 
with $n_n>5$ is named low symmetry cluster (LSC).
 
In the following we use the particle 
diameter $\sigma$ as unit of length and $k_BT$ as unit of energy. 
For the Monte Carlo simulations we use 1000 ``sweeps'' (1000 attempted MC moves 
per particle) as unit of time, for the MD one time unit (``step'') 
corresponds to 27 collisions per particle on average.

The system sizes we simulated were $N=8000$, $14,400$, $64,000$, and $216,000$ 
particles. We studied three densities $N\sigma^3/V = 1.0238$ (volume 
fraction $\eta = 0.5361$), $N\sigma^3/V = 1.0269$ ($\eta = 0.5377$) and 
$N\sigma^3/V=1.03$ ($\eta=0.5393$). 
These densities correspond to chemical potential differences 
between the metastable liquid and the stable, almost completely 
crystalline state of $\Delta \mu \simeq -0.54\;k_BT$, $\Delta \mu \simeq -0.56\;k_BT$ and $\Delta \mu \simeq -0.58\;k_BT$ respectively. 
The interfacial tension is of the order of $0.5\;k_BT/\sigma^2$ 
\cite{Davidchack2000,Davidchack.Morris.Laird:2006}. 
Table \ref{tab:Stats} summarizes the simulation runs we have performed.

Overcompressed liquid configurations were prepared by 
a fast pressure quench from the equilibrated liquid. During the quench we 
monitored crystallinity to ensure that no crystal precursors were formed. 
(As prestructuring during the preparation procedure can have a
significant impact on the nucleation behaviour, we cross-checked the 
quality of our starting configurations: The authors of 
ref.~\cite{Filion:2010} ran trajectories from our starting 
configurations using their simulation code. Within the errorbars we found 
no differences in the crystallization process observed.)

\subsection{Monte Carlo Simulation}
The Monte Carlo simulations were performed at fixed $N$, $V$ and $T$ 
by small translational moves only. 
We let all systems evolve until they crystallized 
and sampled observables every 5,000 sweeps.
Then we prepared movies of the crystalline clusters and played them
backwards in time.  The moment when the 
stable crystallite was reduced to 
a cluster of ca.~10 particles was recorded as ``nucleation time'' $t_n$.
(Apart from the systems of $N=216,000$ particles, 
no system showed more than one crystallization event. 
In the case of $N=216,000$, we used the time when the first crystallite 
formed, as well as the relation ``number of crystllite versus time''
to extract the nucleation rate.)
We also recorded the times when the last particle with $n_b>10$ vanished 
(``last'' when playing the movies backwards) and
the time when the cluster shrunk below 40 particles. For the observables 
we used to extract the the crystallization rates and to discuss the 
crystallisation mechanism, we found no difference between 
these criteria (apart from a slight shift of the time-scale, obviously).
Where times are indicated in the following, each simulation run has been 
shifted by $-t_n$ setting the time to zero at the nucleation event.  

\subsection{Molecular Dynamics}
Molecular dynamics simulations were performed at constant $N$, $V$, and $E$. 
The initial velocities were drawn from a Gaussian distribution and the initial 
mean kinetic energy per particle was set to $3\;k_BT$. The total energy is 
constant over the time of the simulation since all interactions are elastic 
collisions following Newton's equations of motion. In between collisions, 
particles advance ballistically since no force is present. We employed an 
event driven molecular dynamics algorithm 
\cite{Alder1959, Marin1995, Krantz1996, Luba1991}. The analysis was done 
in the same way as for the MC simulation.

\section{Results}

\begin{figure}
  \centering
  \includegraphics[width=0.45\columnwidth]{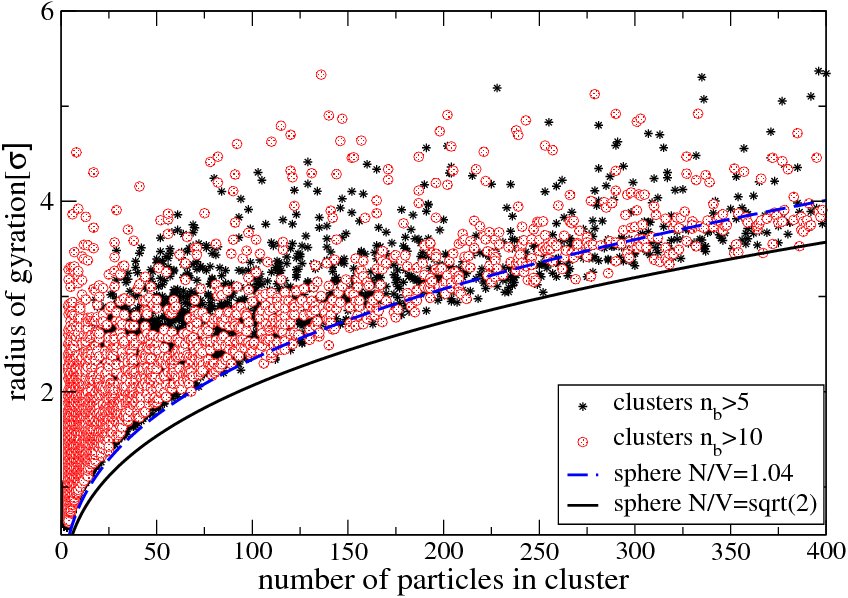}
  \includegraphics[width=0.45\columnwidth]{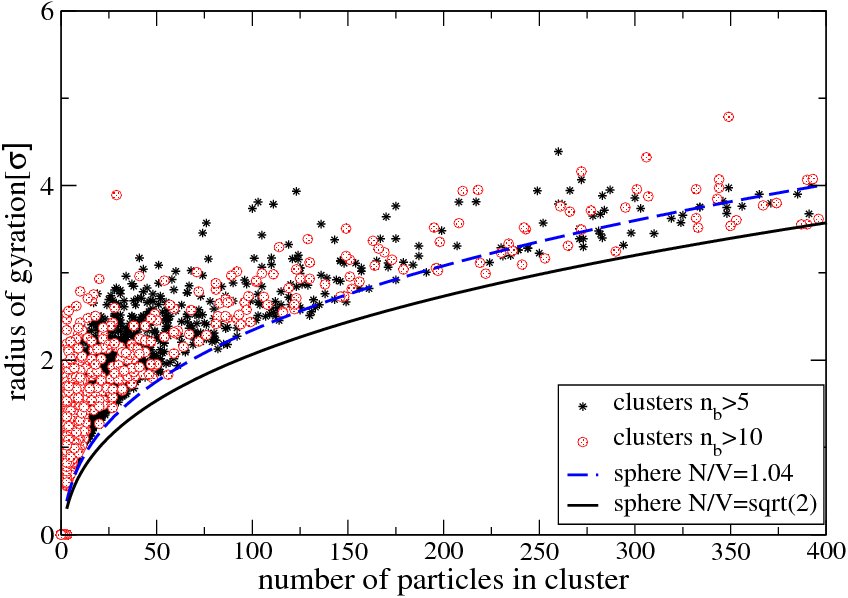}
  \caption{Radius of gyration versus number of particles in cluster for 
    $n_B>5$ (stars) and $n_B>10$ (circles, red online), $N/V=1.03$. 
    Left panel: MC, right panel: MD.
    For comparison $R(N)$
    is plotted for a sphere of $N/V=1.04$ (the density of a hard sphere 
    crystal at coexistence, dashed line, blue online) and a sphere 
    of $N/V = \sqrt{2}$ (solid line).}
  \label{fig:RgN}
\end{figure}

\begin{figure}
  \centering
  \includegraphics[width=0.9\columnwidth]{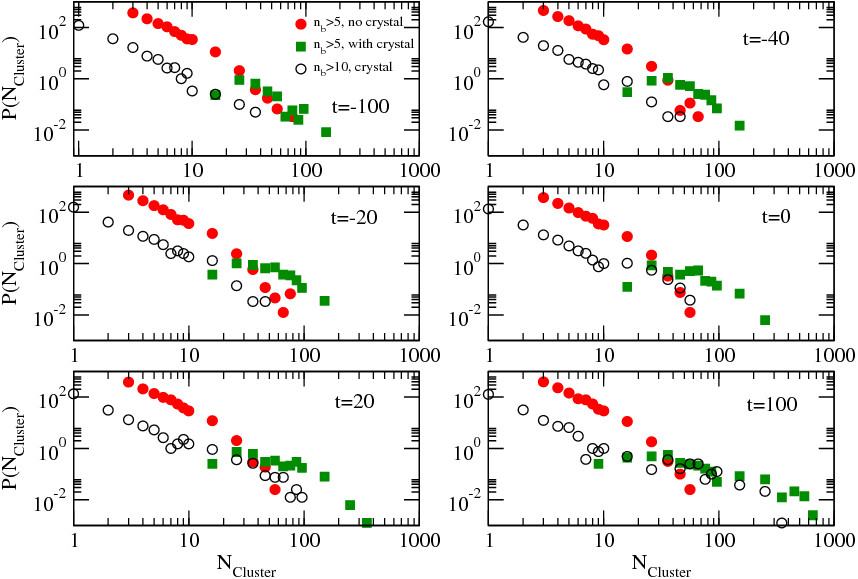}
  \caption{MC: Distribution of cluster sizes
for $n_b>5$ containting no more than 3 particles with $n_b>10$ (filled 
circles, red online), containting more than 3 particles with $n_b>10$ 
(squares, green online) and for $n_b>10$ (open circles). Times are given 
with respect to the time when the first stable crystallite appears (in units 
1000 MC sweeps). $N/V=1.03$.}
  \label{fig:CSD}
\end{figure}

\begin{table}
  \caption{\label{tab:Stats} Simulation details. For the case ${\rm MC^a}$ the rate was computed from the number of clusters in the system as a function of time, while all other rates were obtained from first-nucleation-event-times}
  \begin{ruledtabular}
    \begin{tabular}{lllll}
      type & $N$ & $N\sigma^3/V$  & runs & rate[$\sigma^5/D_l$]\\
      \hline
      MC & 8000 & 1.0269 & 4 & 3.00e-6\\
      MC & 64000 & 1.0269 & 6 & 4.05e-6\\
      MC & 216000 & 1.0269 & 2 & 3.67e-6\\
      MC & 8000 & 1.03 & 8 & 2.10e-5\\
      MC & 14400 & 1.03 & 4 & 2.75e-5\\
      MC & 64000 & 1.03 & 4 & 1.82e-5\\
      MC & 216000 & 1.03 & 5 & 1.07e-5\\
      ${\rm MC^a}$  & 216000 & 1.03 & 5 & 1.73e-5\\
 \hline
      MD & 64000 & 1.0238 & 5 & 7.2e-7\\
      MD & 8000 & 1.0269 & 29 & 1.93e-6\\
      MD & 8000 & 1.03 & 7 & 4.5e-6
     
    \end{tabular}
  \end{ruledtabular}
\end{table}

We first discuss the crystallite structures and then 
present results for the rates.
Fig.~\ref{fig:RgN} shows the radius of gyration $R_g$ versus the number of
particles in a cluster $N_{cluster}$ for all clusters observed in the MC 
simulations (only up to 400 particles in a cluster to keep the graph readable). 
Stars indicate low symmetry clusters, circles crystalline 
clusters. In both, the data from MC and from MD, there is a wide spread 
in $R_g$, structures ranging from 
an almost linear aggregates to very densely packed spheres occur.
Even at 
large crystal sizes (i.e. in the crystal growth regime) there are 
ramified structures. 
Therefore, in the following discussion, we use the 
number of particles in a cluster rather than its 
radius to define a ``cluster size''.

Fig.~\ref{fig:CSD} shows the development of the cluster size distribution
for the MC simulations. 
The data has been averaged over all simulation runs (shifted by the 
``nucleation time'' as explained above). We distinguish between clusters of 
$n_b>5$ with less than 4 particles that have $n_b>10$ (i.e.~''empty'' low 
symmetry clusters that do not contain crystallites, indicated by filled 
circles),
clusters of $n_b>5$ with 4 or more crystalline particles (squares) and
clusters of particles with $n_b>10$ (i.e. crystallites, open circles).
The distribution of empty LSC does not vary with time. 
Just before crystallization sets in, at times $t=-40$ to $t=0$, large low 
symmetry clusters that contain up to several hundred particles are formed, while 
the crystallites (empty circles) are still relatively small. Then the 
crystallites ``follow'' until the two distributions coincide at $t=100$. 
This confirms our previous observation of a precursor mediated 
process \cite{Schilling:2010}.

\begin{figure}
  \centering
  \includegraphics[width=0.9\columnwidth]{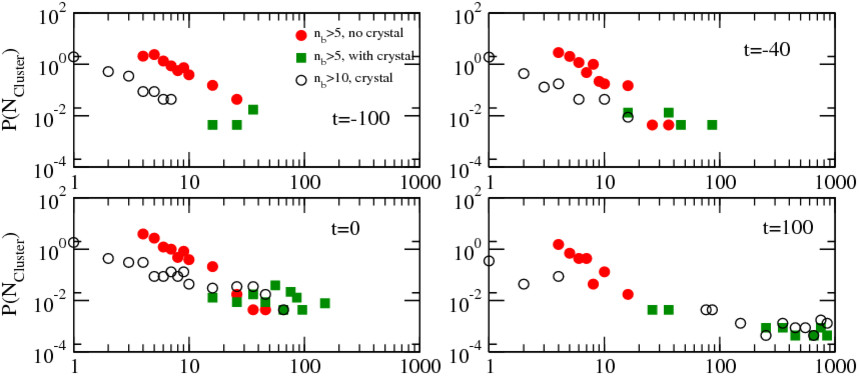}
  \caption{MD: Distribution of cluster sizes for $n_b>5$ containting no more than 3 particles with $n_b>10$ (filled circles, red online), containting more than 3 particles with $n_b>10$ (squares, green online) and for $n_b>10$ (open circles). Times are given with respect to the time when the first stable crystallite appears (in units MD steps). $N/V=1.03$}
  \label{fig:CSDSven}
\end{figure}

Fig.~\ref{fig:CSDSven} shows the cluster size distributions for the MD 
simulation. As in the MC 
case, first low symmetry clusters form, then the crystallites appear. 
Hence the precursor-effect does not depend on the short time dynamics.

\begin{figure}
  \centering
  \includegraphics[width=0.9\columnwidth]{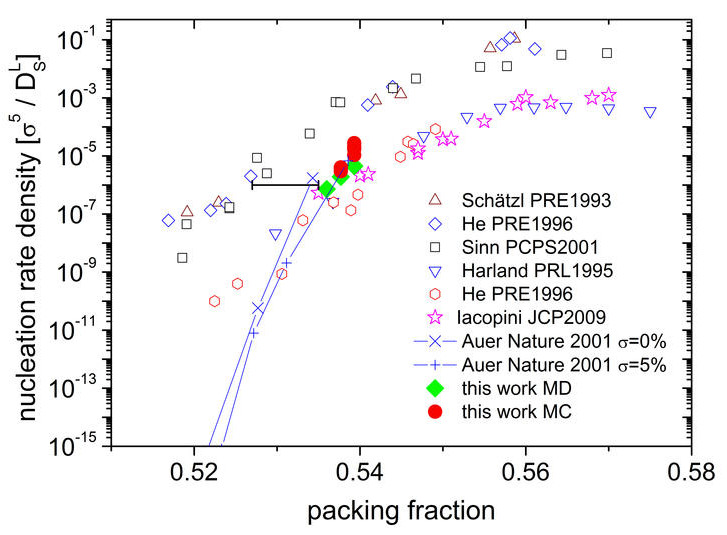}
  \caption{Nucleation rate density versus packing fraction. Solid symbols: data from this work, open symbols and crosses: experiments and simulations as cited and commented on in the main text}
  \label{fig:rateAll}
\end{figure}

In fig.~\ref{fig:rateAll} we compare of the dimensionless 
nucleation rate density with experimental results and results from 
previous simulations. 
We scaled our data with the long-time self-diffusion coefficient $D_l$
extracted from our simulations. The typical error in our data is ca.~$50\%$ 
To scale the experimental data we used the following expression of the long 
time self-diffusion coefficient provided by mode coupling 
theory $D_l/D_0=(1-\Phi/\Phi_{\rm Glass})^v$ using $\Phi_{\rm Glass}=0.58$ and 
$v=2.6$ as determined in experiments. Please note that the data of 
ref.~\cite{Schaetzel:1993, Harland1995, Sinn:2001, He1996} are scaled to the 
freezing volume fraction of monodisperse spheres while the data of 
ref.~\cite{Iacopini2009} is scaled to the freezing volume fraction of 
polydisperse 
spheres with $\sigma=6.5\%$ polydispersity. The typical error determining 
the volume fraction in these experiments is about $\pm 0.004$ as indicated 
by the horizontal error bar while the error in the nucleation rate density 
is about one order of magnitude. Taking the experimental uncertainties into 
account the experimental data sets are not inconsistent. The simulation data 
of Auer and Frenkel \cite{Auer2001} for samples with $5\%$ polydispersity 
have been scaled to the freezing point of monodisperse spheres allowing a 
direct comparison with the older experiments. 
The results obtained in our simulations (solid symbols, 
see also table \ref{tab:Stats}) are in 
good agreement with the experimental data reproducing both their absolute 
values and their slope. They seem to lie below the simulation data 
from ref.~\cite{Auer2001}, but this effect might still be within the error bars.

In summary, we have presented a simulation study of crystallization in 
hard spheres. Both, MD (Newtonian free flight and collisions) and MC 
(quasi-Brownian dynamics) show a precursor mediated crystallization process.
First aggregates of low orientational bond-order form, then crystallites 
grow inside these. The shapes of the crystallites range from ramified 
structures to almost perfectly packed spheres. The crystallisation rates 
agree with the experimental data as well as between MC and MD within 
the errorbars.

\begin{acknowledgements}
We thank Laura Filion for cross-checking start configurations and Martin 
Oettel for fruitful discussions.
This project has been financially supported by the DFG (SFB Tr6 and SPP1296).
\end{acknowledgements}

\end{document}